\documentstyle[12pt,epsfig,rotating]{article}
\def\bcc{\begin{center}} \def\ecc{\end{center}}
\def\beq{\begin{equation}} \def\eeq{\end{equation}}

\def\cl{\centerline}  
   \def\ni{\noindent}
   
\def\la{\langle} \def\ra{\rangle} \def\r#1{$^{[#1]}$}

\begin{document}
\null{}\vskip -2.5cm
\hskip12cm{\large HZPP-9807}

\hskip12cm {\large Dec. 10, 1998}

\vskip2cm
\begin{center}
{\Large SCALING EXPONENTS AND FLUCTUATION STRENGTH 
\vskip0.2cm

IN  HIGH ENERGY COLLISIONS}
\vskip1cm

{\large Liu Lianshou, \ \ Fu Jinghua \ and \ Wu Yuanfang }

 {\small Institute of Particle Physics, Huazhong Normal University, Wuhan
430079 China}
\date{ }
\end{center}

\begin{center}
\begin{minipage}{125mm}
\vskip 0.5in
\begin{center}{\Large ABSTRACT}\end{center}
{\hskip1cm  
The information on dynamical fluctuations that can be extracted from
the anomalous scaling observed recently in hadron-hadron collision
experiments is discussed in some detail.
A parameter ``effective fluctuation strength'' is proposed to 
estimate the strength of dynamical fluctuations.
The method for extracting its
value from the experimentally observed quantities is given. Some 
examples for the application of this method to real experimental data are
presented.}
\end{minipage}
\end{center}
\vskip 1iN
{\large PACS number: 13.85 Hd
\vskip0.5cm

\ni

\qquad \ \ \ \ \ R\'enyi dimension, \ Dynamical fluctuation}
\newpage

The JACEE experiment in 1983\r{1} and afterwards some accelerator
experiments\r{2,3} indicated that there are dynamical fluctuations beyond
the usual statistical ones in high energy multiparticle final states.
These experimental phenomena had caused high interest in studying the 
anomalous scaling in high energy collisions\r{4}.

Recently, the expected anomalous scaling  has 
been observed successfully in both the data of ${\pi}^+$p and K$^+$p 
collisions at 250 GeV/$c$ from NA22\r{5} and the data of pp collisions at
400 GeV/$c$\r{6} from NA27, under the assumption that the dynamical
fluctuations are anisotropic in longitudinal-transverse planes\r{7}.

It is now natural to ask the question: What is the underlying physics of 
these experimental findings, and/or what are the 
properties of the dynamical fluctuations that 
lead to the experimentally observed anomalous scaling.
This question is the subject of a long term study. In
this letter we want to try a limited task in this direction, i.e. to study
the relation between the observed anomalous scaling exponents and the 
strength of dynamical fluctuations within the framwork of a simple model ---
the random cascading $\alpha$-model.

Firstly, let us recall that what is observed directly in experiments is the
anomalous scaling of normalized factorial moments $F_q$, which under the
assumption of Poisson or Bernoulli type of statistical fluctuations are equal 
to the normalized probability moments\r{8}:
$$F_q={\frac{1}{M}}{\sum\limits_{m=1}^{M}}{\frac{\la {p_m}^q\ra}
{{\la p_m\ra}^q}}.\eqno(1)$$
Here, a phase space region $\Delta$ is divided into $M$ 
sub-cells\footnote{Throughout this paper we use the convention popular in 
the study of intermittency and fractality in high energy physics, i.e.
let $M$ in Eq's.~(1) -- (4) be the number of sub-cells in the division
of the phase space region $\Delta$ of topological dimension $D_T$. This
means, in particular, that for a normal geometrical object the fractal
dimension defined via Eq.(4) is always equal to unity. On the contrary,
in the usual definition of fractal dimension\r{9}:
$C_q(l)\propto l^{(q-1)D_q} \ (l \to 0)$, $l$ is the 'edge' (1-dimensional
scale) of the hypercube (sub-cell), and concequently for a normal 
geometrical object $D_q=D_T$.} and $p_m$ is the probability for finding 
particles in the $m$th sub-cell. The anomalous scaling of $F_q$:
$$F_q(M){\propto}M^{\phi_q} \qquad\qquad (M \to \infty)\eqno(2)$$
with non-vanishing indices $\phi_q$, called intermittency (IM) indices,
is an evidence for the existence of dynamical fluctuations\r{8}.

On the other hand, in studying the fractal property of the system, one
usually uses the un-normalized moments\r{9}
$$C_q=\sum\limits_{m=1}^{M} \la {p_m}^q\ra, \eqno(3)$$
which has the anomalous scaling property
  $$C_q(M){\propto}M^{-(q-1)D_q} \qquad\qquad (M \to \infty) . \eqno(4)$$
The indices $D_q$, called multifractal R\'enyi dimensions\r{9}, are related 
to the IM indices $\phi_q$ as
  $$D_q=1-{\frac{\phi_q}{q-1}}. \eqno(5)$$

The question in front of us is as the following: When we obtain successfully
a strict scaling of normalized factorial moments through a proper way of
phase space division, as for example in the cases of Ref.'s [5] and [6], we 
are ready to get the slopes of the ln$F_q\sim$ln$M$ plots, which gives the IM
indices and R\'enyi dimensions. Then, how to extract a characteristic quantity
describing the strength of dynamical fluctuations from the values of these
indices and dimensions. This is the problem we are about to discuss.

It should be noticed that an anomalous power law and a fractal property of a 
system both result from the dynamical fluctuations in this system, and 
therefore the IM indices $\phi_q$
and multifractal R\'enyi dimensions $D_q$ are related to the strength of
dynamical fluctuations. They have, however, their own physical meaning and 
cannot be taken directly as a measure of the dynamical-fluctuation
strength. For example\r{9}, the first-order R\'enyi dimension $D_1$, or 
equivalently the information dimension $D_I$, determines the scaling property 
of the number of boxes containing the dominant part of information; 
the second-order R\'enyi dimension $D_2$, sometimes called correlation 
dimension $\mu$,  
measures the scaling properties of two particle correlations, etc. 
These scaling properties come from a common origin --- the dynamical 
fluctuations. All of them are related to these 
fluctuations but none of them can serve as an appropriate quantity 
for describing the strength of these fluctuations directly.

In order to show the relation between the strength of dynamical fluctuations
and the values of IM indices or R\'enyi dimensions,
let us consider a simple example --- the random 
cascading $\alpha$-model\r{8,10}.  This model describes each multiparticle 
event as a series of steps, in which the initial phase space region 
$\Delta$ is repeatedly divided into $\lambda=2$ parts.        
After $\nu$ steps we get $M= {2^ \nu}$ sub-cells of size
$\delta = {\Delta}/{M}$.
At each step $\nu$ the normalized particle density is obtained in
each of the two parts by multiplication of the normalized density
in the step $\nu-1$ by a particular value of the random variable
${\omega}_{\nu{ j_ \nu}}$, where
$j_ \nu$ is the position of a sub-cell at the ${\nu}$th step
(1$ \leq$ $j_{\nu}$ $\leq$ ${2^\nu}$).

The elementary fluctuation probability $\omega$ can be chosen in various 
ways\r{8,10}. The simplest way that provides a characteristic 
parameter for describing the fluctuation strength is to choose it as\r{10}
   $${\omega}_{\nu,2j-1}={\frac{1}{2}}(1+{\alpha}r) \ \ \ \  ;  \ \ \ \
{\omega}_{\nu,2j}={\frac{1}{2}}(1-{\alpha}r), \eqno(6)$$
in which $r$ is a random number distributed uniformly in the interval
$[-1,1]$, $j$ is an integer ($1\leq j \leq 2^{\nu-1}$),
$\alpha$ is a positive constant taking value in the range [0,1],
    $$ 0 \leq \alpha \leq 1\ . \eqno(7) $$
The value of $\alpha$ determines the possible region of $\omega$,
    $${\frac{1-\alpha}{2}}<{\omega}<{\frac{1+\alpha}{2}}. $$

Let us note that $\omega$ defines the way how particles are distributed from
step to step between the two pieces of a given cell, i.e. it 
characterizes the strength of multiplicity fluctuations in cell-division, 
and $\alpha$ determines the width of the possible values of $\omega$,
therefore, ${\alpha}$ is the characteristic
quantity describing the {\it strength of dynamical fluctuations} in this 
version of random cascading model. In the following, we will analyse
the relation between the strength parameter ${\alpha}$ and the
multifractal dimensions $D_q$ in this model.

In the random cascading $\alpha$-model
$$F_q(M)={\frac{\la\omega ^q(1) \cdots \omega ^q(\nu)\ra}
{\la\omega\ra^{q \nu}}} . \eqno(8)$$
In the limit of large $\nu$, the distribution of random variable
$\zeta ={\sum\limits_{i=1}^{\nu}}\ln \omega (i)$ approaches a Gaussian\r{8}
$$p(\zeta)d \zeta=(2 \pi \nu)^{-1/2} \sigma ^{-1}\exp[-(\zeta-\nu 
\overline{V})^2/2\nu {\sigma}^2]d\zeta$$ 
with
$$\sigma ^2=\int P(\omega)(\ln (\omega)-\overline{V})^2d \omega
\ , \qquad \overline{V}=\int P(\omega) \ln (\omega)d\omega . \eqno(9)$$
Explicit calculation of multifractal dimensions in this limit 
gives
$$D_q=1-{\frac{1}{2\ln\lambda}}{\sigma ^2}q . \eqno(10) $$

A characteristic feature of the Gaussian approximation is the proportionality
of $1-D_q$ and $q$ as can be seen from Eq.(10):  
$$ \frac{1-D_q}{q}= {\frac{1}{2\ln 2}} \sigma^2. $$
Here and in the following we take $\lambda=2$ for simplicity.
In Fig.1 is shown the relation between $1-D_q$ and $q$
for $\alpha=0.1$ to 0.5. It can be seen from the figure that when $\alpha$ is 
small there is a fairly good linear relation between $1-D_q$ and $q$, so that
the Gaussian approximation is sufficiently good in these cases.

To get a relation between $D_q$ and $\alpha$, we calculate the variance 
$\sigma^2$, appearing in eq.(10), of the random variable ln$\omega$ 
$$\sigma ^2  =  \la\ln ^2 \omega \ra- \la\ln \omega\ra^2
 = {\frac{1}{3}}{\alpha ^2}+{\frac{2}{3}}{\alpha ^4}+\cdots\ \ . \eqno(11)$$
Under linear approximation
$$ \sigma ^2  \approx {\frac{1}{3}}{\alpha ^2} . \eqno(12) $$

How ${\sigma}$ is related to ${\alpha}$ is shown in Fig.2. In this figure,
the full circles represent the result without linear approximation, the
dashed line indicates the linear-approximation. When $\alpha$ is not
very large ($\alpha$ $\leq$ 0.5 say), the two results are nearly equal. 
Therefore, both Gaussian and linear approximations can be used when
$\alpha$  $\leq$ 0.5. This region of $\alpha$ is sufficient for the limited 
range available in actual experiments, cf. Table~I.  

Substituting Eq.(12) into Eq.(10), we get
$$D_q=1-{\frac{1}{6 \ln 2}}q \alpha ^2$$
or
$$ \alpha = \sqrt{\frac{6 \ln 2}{q}(1-D_q)} . \eqno(13) $$
It can be seen from Eq.(13) that, in the random cascading $\alpha$-model, with 
small $\alpha$, the strength parameter $\alpha$ of dynamical fluctuations is 
related to the multifractal dimensions $D_q$ by a very simple relation. Using 
Eq.(13) we can get an approximate value of the fluctuation parameter $\alpha$ 
as long as the multifractal dimension of any order $q$ is known. In high 
energy experiments, the second-order multifractal dimension $D_2$ is most easy 
to obtain. For this reason, the r.h.s. of Eq.(13) for $q=2$,
$\sqrt{3\ln2(1-D_2)} \approx \sqrt{2(1-D_2)}$, can be taken as
a characteristic quantity for the strength of dynamical fluctuations.

In Fig.3 the values of 
$\sqrt{2(1-D_2)}$, denoted by $\alpha_{\rm eff}$ (cf. Eq.(14)), are ploted  
against the model parameter $\alpha$ varying from 0 to 1. The dashed line
corresponds to $\alpha_{\rm eff} = \alpha$. 
From the figure we can see that $\sqrt{2(1-D_2)}$ has almost equal
value with $\alpha$, especially when $\alpha$ is not very large. 
This means that $\sqrt{2(1-D_2)}$ represents the value of $\alpha$ fairly well. 
Thus, we have been successful in obtaining an estimation of the
fluctuation strength $\alpha$ in terms of the second-order fractal 
dimension $D_2$ in the framework of the random cascading $\alpha$-model.

The above results are obtained from a special model.
In the general case, when the underlying dynamics is unclear, it is hard
to define the ``strength of dynamical fluctuations'' strictly. 
In these cases, we can make use of the results obtained from the
random cascading $\alpha$-model as an estimation for this strength.

Thus, for an arbitrary process that has anomalous scaling property, we 
define an {\it effective fluctuation strength} 
  $$ \alpha_{\rm eff}=  \sqrt{2(1-D_2)} = \sqrt{2\phi_2} .\eqno(14) $$
as an estimation of the strength of the dynamical fluctuations taking 
place in this process.  Its physical meaning is:
\medskip

\begin{center}
\begin{minipage}{145mm}
\qquad{\it The effective fluctuation strength $\alpha_{\rm eff}$  of an 
arbitrary process is
the fluctuation strength of a random cascading $\alpha$-model 
with elementary partition number $\lambda=2$ 
that can give the same value of second-order IM index $\phi_2$ 
(within the Gaussian and linear approximation) as this process.  }
\end{minipage}
\end{center} 
In getting the last equality of Eq.(14) the relation (5) between the IM 
indices and R\'enyi dimensions has been used.

Using the effective fluctuation strength defined above, we are now able to
compare the strength of dynamical fluctuations in different collision 
processes. In the following we will give some examples. Before doing
that, a question has still to be considered. 

In real experiments the dynamical fluctuations exist in 
higher-dimension\r{11} and are usually anisotropic\r{5-7} with a particular 
value of Hurst exponent $H_{\parallel\perp}$. How can we use the definition 
(14) of effective fluctuation strength, which depends on a one-dimensional 
$\alpha$-model with elementary partition number $\lambda=2$, to these cases?

In answering this question, let us note that in case of anisotropic
dynamical fluctuation (self-affine fractal), the partition numbers in
different phase space directions cannot be simultaneously equal to
integer values. The method of factorial moment analysis with non-integer
partition\r{12,13} has to be used
and the resulting $F_2^{\rm 3D}$ for arbitrary value of $M^{\rm 3D}$
all lie on a same straight line. Therefore, we can freely choose
$M^{\rm 3D}=2^\nu, \nu=1,2,\dots$. The anomalous scaling property of
such a 3-D fractal is equivalent to that of a 1-D fractal with elementary 
partition number $\lambda=2$.
Therefore, our definition (14) for effective fluctuation strength is
applicable also to this case.

Now, let us turn to the examples for the application of 
effective fluctuation strength to real experimental data. For this purpose
we have to choose those data that possess good anmalous scaling property.

For hadron-hadron collisions, the presently available 3-D data that have good
scaling property are the self-affinely analysed data for ${\pi}^+$p and 
K$^+$p collisions at 250 GeV/$c$ from NA22\r{5} with Hurst exponents
 $ H_{yp_{\rm t}}= H_{y\varphi}=0.475 \ \ ,\ \ H_{p_{\rm t}\varphi}=1 .  $
The second-order IM index is obtained as\r{13} : \ 
$\phi_2^{\rm 3D}=0.061 \pm 0.010$.

Another example of hadron-hadron collisions is the
pp collisions at 400 GeV/$c$ from NA27\r{6}. In this case, only
2-D ($\eta,\varphi$) data are available due to lack of momentum 
measurement. The Hurst exponent in the ($\eta,\varphi$) plane is found to be 
$H_{\eta\varphi}=0.74$. Fitting the results to  a straight line gives 
$\phi_2^{\rm 2D}=0.051 \pm 0.004$. 

As an example of e$^+$e$^-$ collisions we take the data from DELPHI\r{14}. 
After omitting the first point to eliminate the influence of momentum 
conservation\r{15}, a good fit to a straight line comes out\footnote{The 
reason why the e$^+$e$^-$ data have good scaling property already for the 
Hurst exponent $H=1$ will be discussed elsewhere.}, cf. Fig.4. The 
second-order IM index is then obtained as $\phi_2^{\rm 3D}= 0.099 \pm 0.005 $. 

The resulting effective fluctuation strengths $\alpha_{\rm eff}$ for these 
three cases are listed in the last column of Table I. 
Its physical meaning is that, the anomalous 
scaling of the second-order factorial moments in these 3 experiments can be 
produced by random cascading processes having fluctuation strength
$\alpha \approx 0.349, 0.319, 0.446$, respectively. 

\begin{center}
Table I \ \ The Hurst exponents, IM indices, second order R\'enyi dimensions 
and effective fluctuation strengths for the data from 3 experiments

 \begin{tabular}{|l|c|c|c|c|}\hline
Experiment & Hurst exponent & $\phi_2$ & $D_2$ & $\alpha_{\rm eff}$ \\  \hline
NA22 \ (3D) & $H_{y\perp}=0.475, H_{p_{\rm t}\varphi}=1$
& 0.061$\pm$0.010 & 0.939$\pm$0.010 
& 0.349$\pm$0.028\\ \hline
NA27 \ (2D) & $H_{\eta\varphi}=0.74$ & 0.051$\pm$0.004& 0.949$\pm$0.004
& 0.319$\pm$0.014\\ \hline
DELPHI (3D) & $H=1$ & 0.099$\pm$0.005 & 0.901$\pm$0.005 & 
0.446$\pm$0.012  \\ \hline
 \end{tabular} \end{center}

Let us notice that all the values of  $\phi_2$ and $D_2$ 
listed in Table I are near by the boundaries of their allowed range,
i.e. 0 for $\phi_2$ and 1 for $D_2$. This would give us an impression 
that the strength of dynamical fluctuations in these experiments are all
marginal.  This is, however, wrong.  From the last column 
of Table I we can see that the values of $\alpha_{\rm eff}$ 
for hadron-hadron collisions are approximately equal to 1/3 while that for
e$^+$e$^-$ collisions is close to 1/2. Since the allowed range of the 
parameter $\alpha$ characterizing dynamical fluctuations is [0, 1], cf. Eq.(7),
the above results show that the dynamical fluctuations are nearly 
equal to one third of the maximum possible strength in hadron-hadron collisions 
and about half of the maximum possible  strength in e$^+$e$^-$ case.  This
gives us, at least qualitatively, a feeling about the strength 
of dynamical fluctuations in these collision processes.
 
\bigskip
\noindent{Acknowledgements}

We are grateful to Zhang Yang and Liu Fuming for helpful discussions. 
This work is supported in part by the National Natural Science Fund of China. 

\vfill

\newpage

\null{}
\vskip2cm

\begin{picture}(260,240)
\put(-125,-180)    
{\epsfig{file=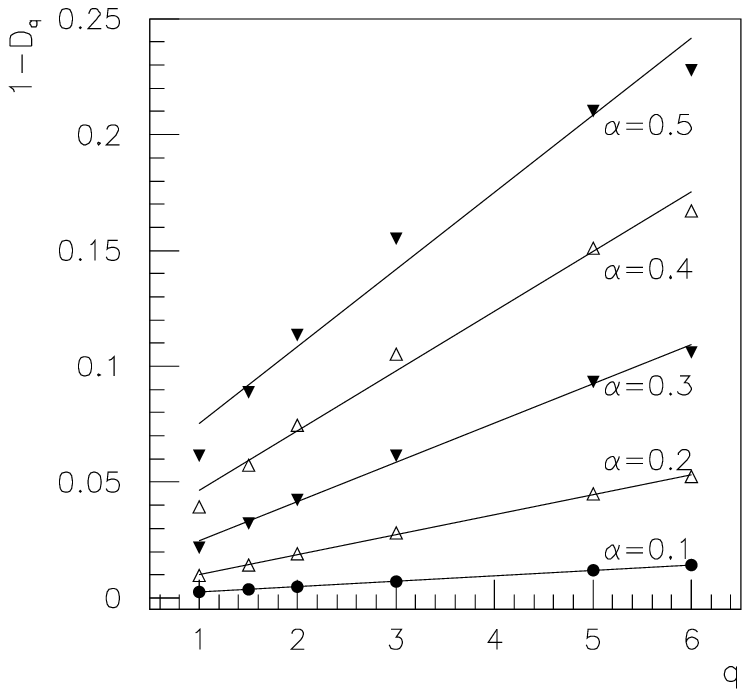,bbllx=0cm,bblly=0cm,
	   bburx=8cm,bbury=6cm}}
\end{picture}

\vskip-3.5cm
 \cl{ Fig.~1 \ Relation between the $1-D_q$ and $q$ for 5 values of
model parameter  $\alpha$.}


\begin{picture}(260,240)
\put(-120,-360)   
{\epsfig{file=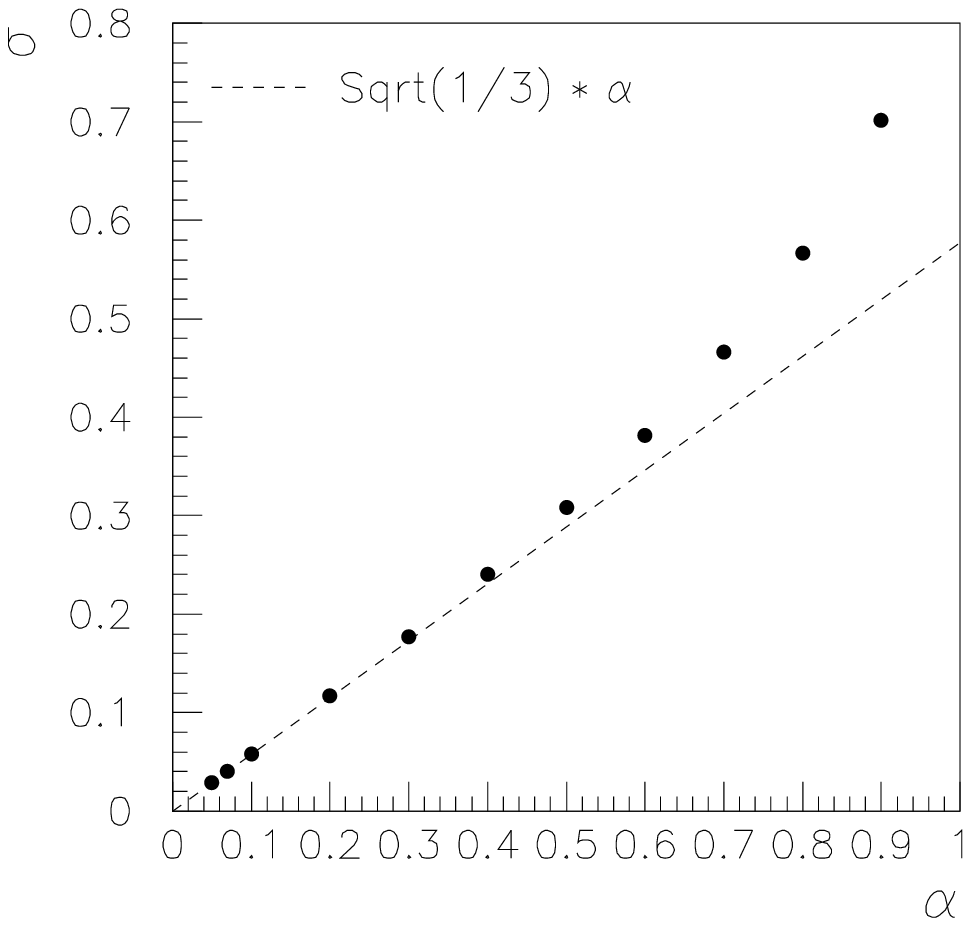,bbllx=0cm,bblly=0cm,
	   bburx=8cm,bbury=6cm}}
\end{picture}

\vskip3.5cm
 \cl{ Fig.~2 \ Relation between the standard deviation  
$\sigma$ of random variable ln$\omega$}

\hskip1.6cm{ and model parameter~$\alpha$.}

\newpage

\null{}
\vskip2cm

\begin{picture}(260,240)
\put(-50,-300)
{\epsfig{file=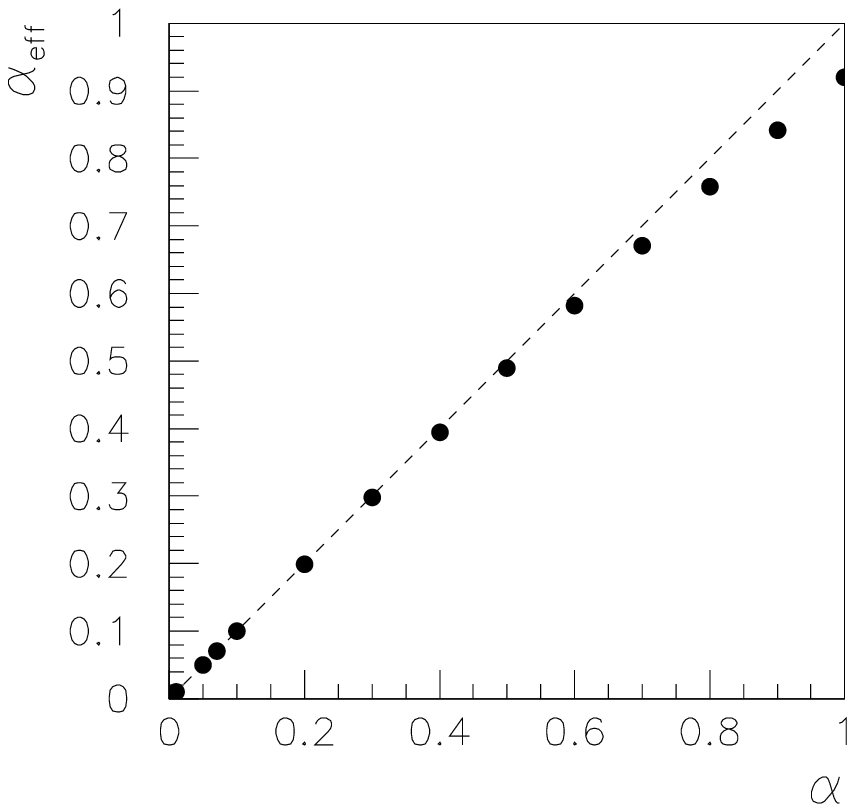,bbllx=0cm,bblly=0cm,
	   bburx=8cm,bbury=6cm}}
\end{picture}

\vskip-2.8cm
 \cl{ Fig.~3 \  Relation between the effective fluctuation 
strength $\alpha_{\rm eff}$ and model parameter $\alpha$. }

\qquad The dashed line corresponds to $\alpha_{\rm eff} = \alpha$ 

\begin{picture}(260,240)
\put(-120,-320)
{\epsfig{file=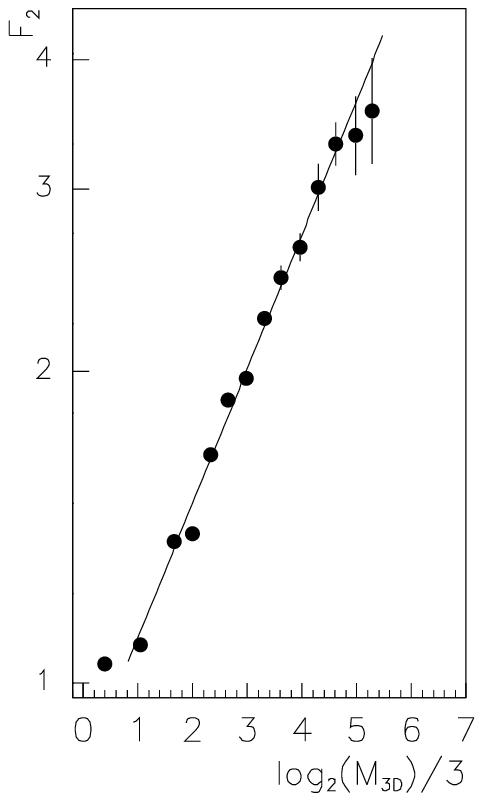,bbllx=0cm,bblly=0cm,
	   bburx=8cm,bbury=6cm}}
\end{picture}

\vskip2.2cm
 \cl{ Fig.~4 \ The anomalous scaling of second order 3-D factorial moment 
of e$^+$e$^-$ collisions }

\hskip1cm
at 91 GeV (data taken from Ref.[14]).

\end{document}